\documentclass{elsarticle}

\usepackage{amsmath}
\usepackage{graphicx}
\usepackage{multirow}
\usepackage{multicol}

\DeclareGraphicsExtensions{.eps} 

\usepackage{textcomp} 

\usepackage[
pdfauthor={Olivier Bourrion},
pdftitle={UCTM: A User friendly Configurable Trigger, scaler and delay Module for nuclear and particle physics},
pdfsubject={UCTM},
pdfkeywords={},
colorlinks={true},
linkcolor=red,
citecolor=green,
urlcolor=cyan,
pdfstartview={FitH}
]{hyperref}

\begin{document}
\begin{frontmatter}
 
\title{UCTM: A User friendly Configurable Trigger, scaler and delay Module for nuclear  and particle physics}
\author[LPSC]{O.~Bourrion\corref{cor1}}
\ead{olivier.bourrion@lpsc.in2p3.fr}
\author[LPSC]{B.~Boyer}
\author[LPSC]{L.~Derome}

\cortext[cor1]{Corresponding author}
\address[LPSC]{Laboratoire de Physique Subatomique et de Cosmologie,\\ 
Universit\'e Joseph Fourier Grenoble 1,\\
  CNRS/IN2P3, Institut Polytechnique de Grenoble,\\
  53, rue des Martyrs, Grenoble, France}


\begin{abstract}
A configurable trigger scaler and delay NIM module has been designed to equip nuclear physics experiments and lab teaching classes.  It is configurable through a Graphical User Interface (GUI) and provides a large number of possible trigger conditions without any Hardware Description Language (HDL) required knowledge.
The  design, performances and typical applications are presented.

\end{abstract}

\end{frontmatter}
\section{Introduction}
\begin{figure}[ht]
\begin{center}
\includegraphics[width=0.7\textwidth]{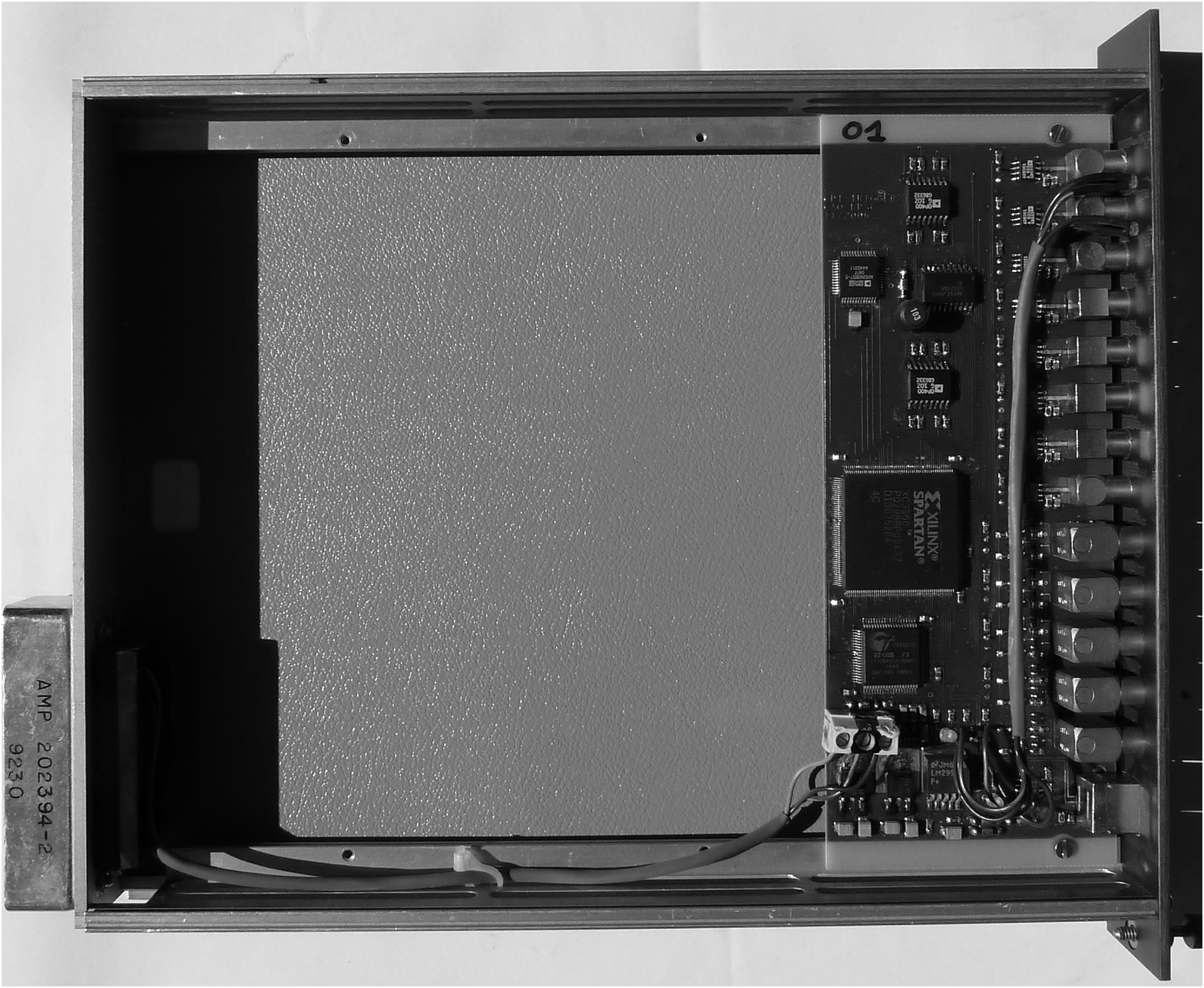}
\hspace{1cm}
\includegraphics[width=1.2cm]{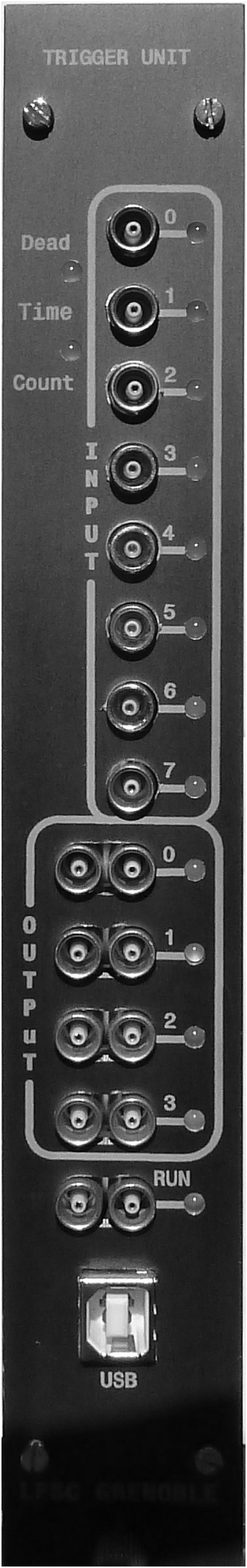}
\caption{Picture of the electronic board inserted in the module (left hand
side) and of the front panel (right hand side)}
\label{photocarte}
\end{center}
\end{figure}

For small nuclear experiments or for teaching in lab classes, it is often necessary to be able to quickly setup an instrumentation bench. The setup usually features logic modules (discriminators, AND gates, OR gates, ...) and delay modules for trigger building as well as scalers. Therefore, as a generic building block, a User Configurable Trigger scaler and delay NIM Module (UCTM) has been designed with the objective of being usable by physicists or students having no particular knowledge in any Hardware Description Language (HDL). The module, which has eight analog inputs, can provide four configurable digital outputs. The solution is composed of two interdependent parts. The first part is a configurable electronics board relying on a FPGA for the digital functions. Fig.~\ref{photocarte} shows a picture of the electronics board inserted in the module (left hand side) and of the front panel (right hand side).

\begin{figure}[ht]
\begin{center}
\includegraphics[width=0.9\textwidth]{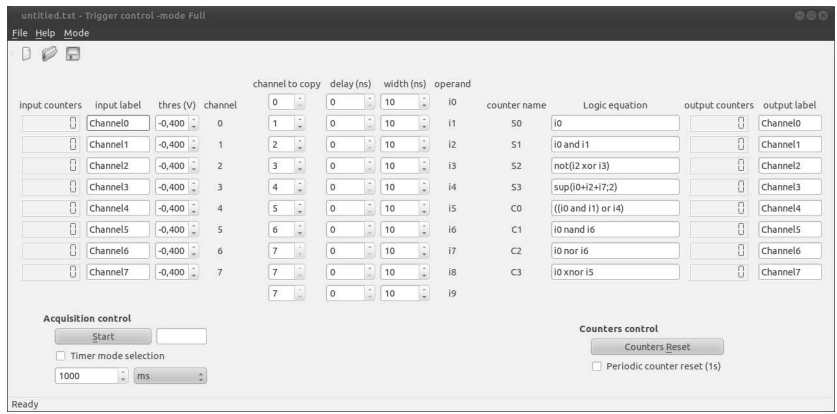}
\caption{Screen capture of the GUI. The upper panel is divided in three parts.
On its l.h.s appears the input scalers, the threshold tuning and their user
selectable labels. The middle part is used to select the duplication/routing
and to enter the delaying/shaping settings. Finally, on its r.h.s 8 trigger
equations, using 1 up to 10 operands, can be entered with their associated
label and activity counters. The lower part of the panel contains the
acquisition and the counter controls.}
\label{GUI_UCTM_NB}
\end{center}
\end{figure}

The second part is a Graphical User Interface (GUI) allowing scalers readout and easy trigger/delay configuration. 
The trigger equations, while permitting any combinatorial operation (with up to ten operands), are directly entered in the GUI before usage. Trigger equation examples can be seen in fig.~\ref{GUI_UCTM_NB}. 

At each run start up, the equations are used to compute each output truth table which are loaded in the FPGA memory blocks. Note that consequently the FPGA firmware is never modified, so there is no need for synthesis and  placer-router tools to configure the board, thus avoiding any tool licensing issue.

This paper is organized as follows: section~\ref{HardwareSec} presents the hardware design, section~\ref{FPGASec} describes the FPGA contents. The control and readout software is presented in section~\ref{SoftSec}. Eventually, two typical example applications are described in section~\ref{ExampleApplications} and a short summary is given in section~\ref{SummarySec}.

\section{Hardware development}
\label{HardwareSec}
\begin{figure}[ht]
\begin{center}
\includegraphics[width=0.85\textwidth]{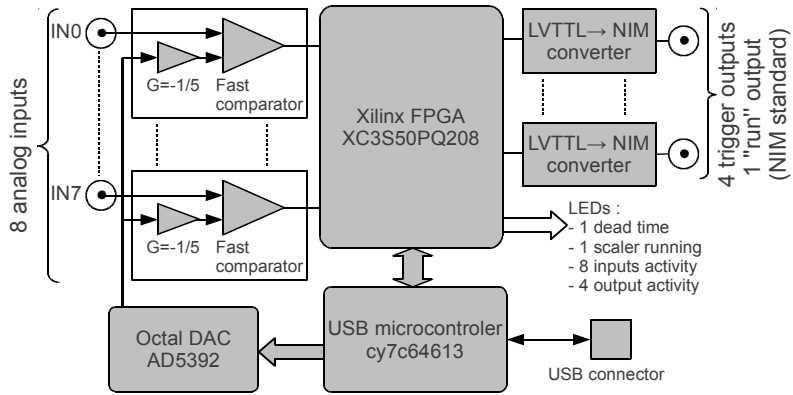}
\caption{Descriptive block diagram of the electronics. Each one of the eight inputs feeds a fast comparator having a tunable threshold. The outputs of the comparators feed the FPGA which hosts the triggering, the delaying/shaping functionality and the counters. The electronics provide 4 duplicated trigger outputs and one ``run'' output in the NIM standard. LEDs are used to provide visual information about: input/output activity, dead time and running mode. A USB micro-controller is used four communication and FPGA configuration.}
\label{UCTMBoardDiag}
\end{center}
\end{figure}

The descriptive block diagram of the electronics is shown in fig.~\ref{UCTMBoardDiag}. Each one of the eight inputs feeds a fast comparator whose threshold is set by a 12 bit DAC. The 
tuning range lies between -1\,V and 0\,V and has a step resolution of 60\,\textmu V. The outputs of the comparators feed the FPGA which hosts the triggering, the delaying/shaping functionality and the counters. Note that the chosen FPGA (Xilinx XC3S50PQ208) contains sufficient amount of user logic and memory block to fulfill the functional requirement (details in section~\ref{FPGASec}). 

The electronics provide 4 duplicated trigger outputs and one ``run'' output in the still widely used NIM standard. The latter indicates when the scalers are counting. Each input/output is associated with a LED to inform visually whether it is toggling or not. Two more LEDs are used to indicate the ``run'' status and the dead time.

A USB micro-controller is used to communicate with the FPGA and DAC, and thus to perform slow control and readout. The micro-controller, which is soft loaded at startup, is also used to configure the FPGA. This feature makes unnecessary for the electronics to hold a FPGA dedicated configuration memory and allows easy firmware upgrade.

\begin{figure}[ht]
\begin{center}
\includegraphics[width=0.6\textwidth]{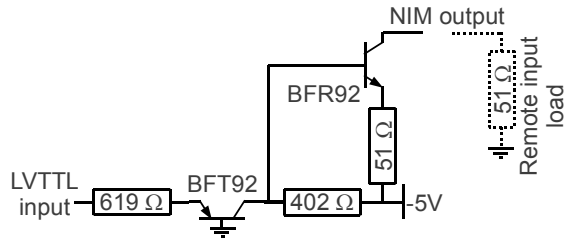}
\caption{Schematic of the LVTTL to NIM converter, using UHF transistors to achieve the minimal possible propagation time (about 2.8\,ns).}
\label{TTL2NIM}
\end{center}
\end{figure}

The TTL to NIM conversion is done with a discrete circuit shown fig.~\ref{TTL2NIM}. It uses Ultra High Frequency (UHF) transistors to achieve the minimal possible propagation time (about 2.8\,ns). The chosen input resistor is high enough to allow proper operation of the FPGA output (low output pad current draw) and the resistors ratio is chosen to be compatible with a 3.3\,V TTL standard (LVTTL).

\section{FPGA firmware description}
\label{FPGASec}
\begin{figure}[ht]
\begin{center}
\includegraphics[width=0.8\textwidth]{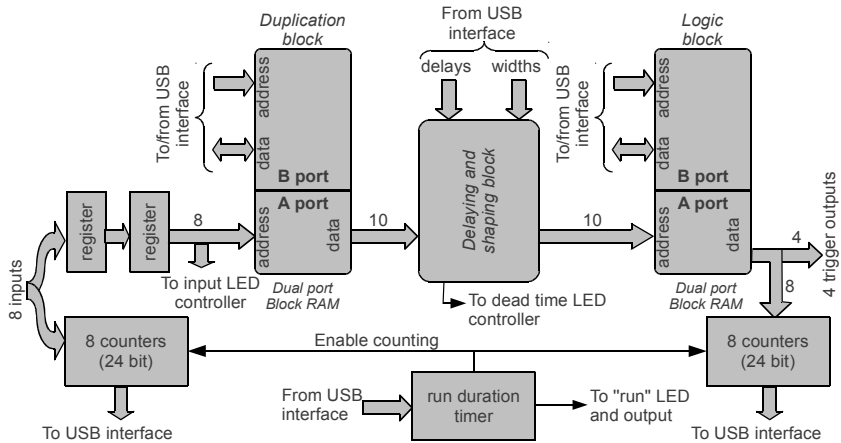}
\caption{Diagram of the FPGA content. The main firmware part are depicted,
i.e. the input synchronization registers, the duplication block, the
delay/shaping block and the logic block. The LED controllers and the USB
micro-controller interface used to configure the blocks are not shown. The
trigger path is synchronous and clocked at 100\,MHz.}
\label{UCTM_Firmware}
\end{center}
\end{figure}

In the diagram shown in fig.~\ref{UCTM_Firmware}, the main firmware parts are depicted, i.e. the input synchronization registers, the duplication block, the delay/shaping block and the logic block. The LED controllers and the USB micro-controller interface used to configure the blocks are not shown. 

Each input is used to supply a synchronization register on one side and a 24
bit counter on the other side. The trigger path, from synchronization to output
is clocked at 100\,MHz. This clocking eases the FPGA design implementation, but
induces an added jitter of 10\,ns.

The ``duplication block'', which is equivalent to a fanout buffer, uses eight
inputs and can provide up to ten signals to the following blocks. It is used to
replicate a chosen input a preprogrammed number of times. This function is
achieved by using the concatenated input signals as an eight bit address  which
points to a ten bit word containing the duplication result. This $2^8 \times
10$ bits duplication matrix is computed beforehand for all input combination to
eventually associate each address bit to one or several data output. 

The ``delay/shaping'' block is used to delay the signal from 0 to
\mbox{2\textsuperscript{13}-1} clock cycles and to adjust its width within a
range of 1 to \mbox{2\textsuperscript{13}-1} clock cycles. This function is
performed by two adjustable monostables (one for delaying and one for 
width adjustment). Provided that the input signal edge is used as the reference
time to perform the delaying and shaping, a signal width can be reduced as well
as enlarged. Any new input pulse received by a delaying/shaping block while the
previous processing is still in progress is ignored. Consequently, a
dead time signal is provided at the block output and used to flash a dead time
LED whenever any of the channel is busy. 

The last and most important block of the firmware is the ``logic block''
which behaves in a similar way as a Look Up Table (LUT) in any FPGA. Similarly
to the ``duplication block'', the ``logic block'' is preloaded with a truth
table giving the expected output vector as a function of the input vector. The
input vector, composed of the ``delay/shaping block'' output is used for the
memory address and the memory data is used as the output vector.
Hence, each memory output bit is a function of up to ten inputs (or operands)
and the latency of this block is independent of the trigger equation
complexity. In a simplified example shown in  table~\ref{ORtable}, a four
words deep memory block is used to implement a two inputs OR gate. The memory
block data output directly reflects the expected behavior when using the two
input bit as the address vector and by conveniently preloading the memory content
with the OR gate truth table.

The trigger output rates are monitored by 24 bit counters. Out of the eight
trigger outputs, only four are actually routed outside of the FPGA, the
remaining four are feeding the internal counters only.

\begin{table}
\begin{center}
\begin{tabular}{|c|c|c|c|}
  \multicolumn{1}{c}{input vector} & \multicolumn{1}{c}{address} & \multicolumn{1}{c}{output vector}\\
\hline
  0 0 & 0 & 0\\  
\hline
  0 1 & 1 &  1\\  
\hline
  1 0 & 2 &  1 \\  
\hline
  1 1 & 3 &  1\\  
\hline
\end{tabular}
\end{center}
\caption{Example usage of a four words deep memory block to implement an OR gate. By conveniently preloading the memory content and by using the 2 input bit as address, the data output reflects the expected behavior.}
\label{ORtable}
\end{table} 

The input and output counters are directly using their inputs as clock signal, thus they can monitor rates higher than the system clock, i.e 100\,MHz. All activity counters can be free running or gated by a duration timer which can be adjusted with a step resolution of 1\,ms from 0 (not used) to 2\textsuperscript{32}-1 (about 50 days).

The minimal propagation latency in the FPGA (delays set to zero) due to this design is four clock cycles, which is equivalent to 40\,ns. Accounting for all the electronics delays, i.e. fast comparator, FPGA input/output delays and TTL to LVTTL NIM conversion, the total minimal board latency is less than 70\,ns.

\section{Readout and control software}
\label{SoftSec}
The readout and control software, which is written in C++, is composed of two
layers: the Application Programming Interface (API) and the Graphical User
Interface (GUI). The API is using open source software drivers to control the
USB port \cite{libusb,libwinusb}. Aside from providing the basic functionality
for accessing the electronics memory map, it also provides the trigger equation
parser, the truth table and duplication matrix generators.

The main building block of the API is the equation interpreter and computer. It
is designed to interpret equations with up to ten operands and to manage all the
basic binary operators, such as AND, OR, XOR (exclusive OR), XNOR (complemented
exclusive OR), NOR and NAND, as well as the unary operators: NOT() and SUP().
SUP() is the multiplicity operator, accepting a list of operands and the target
multiplicity number. Trigger equation examples can be seen in
fig.~\ref{GUI_UCTM_NB}.

\begin{figure}[ht]
\begin{center}
\includegraphics[width=\textwidth]{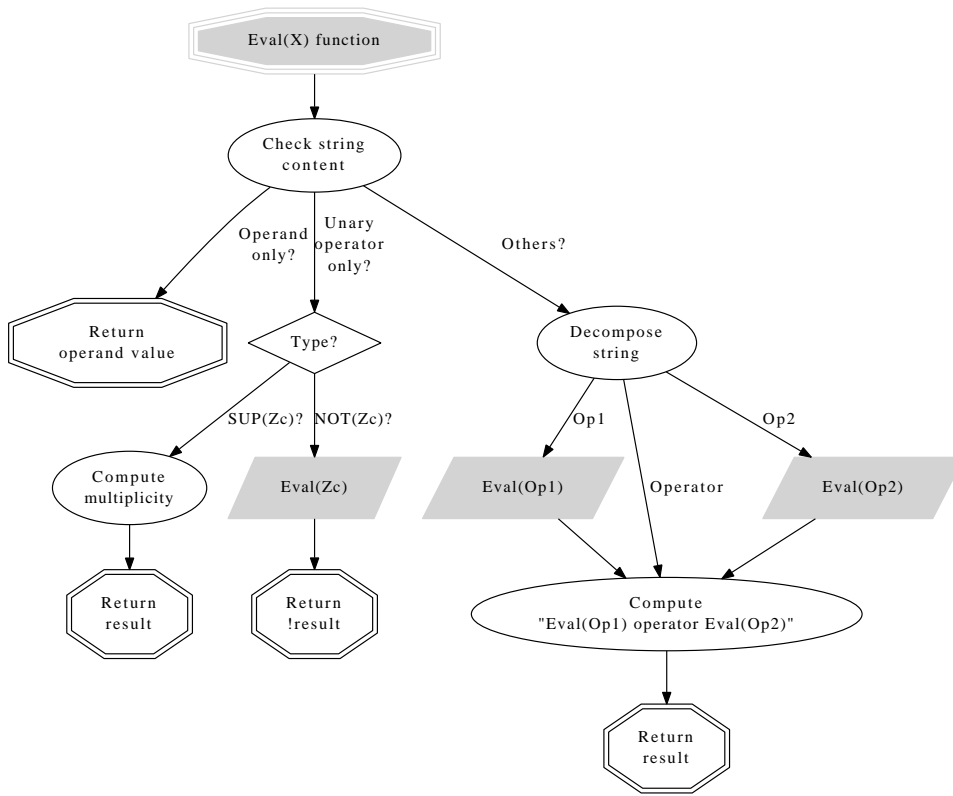}
\caption{Algorithmic view of the logic equation interpreter and computer. The
function uses recursive call to decompose further and further the equation
string in substring until it can finally return either an operand value or a
basic operation result (unary or binary).}
\label{LogicInterpreter}
\end{center}
\end{figure}

The logic calculator operates in two steps. At first, before usage, the formula
string is checked for syntax errors and in a second time it is evaluated against an input
vector.  As depicted in fig.~\ref{LogicInterpreter}, the calculator parses the
equation and uses recursive calls to decompose further and further the equation
string in substrings until it can finally return either an operand value or a
basic operation result (unary or binary).  The truth table of each trigger
equation is built by feeding all possible input vector values to the logic
interpreter. For each identical input condition, the eight computed trigger results
are concatenated in a single byte to form the data to be stored in the memory
block.

To ease the electronics usage, efforts were made to design an intuitive GUI
\cite{Qt}, shown in fig.~\ref{GUI_UCTM_NB}. The upper part of the panel is
divided in three zones. On its l.h.s appears the input scalers, the threshold
tuning and their user selectable labels. The middle part is used to select the
duplication/routing and to enter the delaying/shaping settings. Finally, on its
r.h.s eight trigger equations, using one up to ten operands, can be entered with their
associated labels and activity counters.  The lower part of the panel contains
the acquisition and the counter controls.

\subsection{Module configuration methodology}

To use this trigger module, the user should configure the following elements
through the GUI (from the left to the right of the panel shown
fig.~\ref{GUI_UCTM_NB})~:
\begin{enumerate}
\item Set the threshold value for each used input. Optionally, each of these
input channels can be associated with a text label. 

\item Choose the input channel number that should be used by each
delaying/shaping element in the ``channel to copy'' fields of the GUI. This
corresponds to the replication or rerouting part of the module. 


\item Adjust the delaying and shaping values for each signal. Each value
can be set with a precision of 20\,ns .

\item Enter logic equations (up to eight), optionally each equation can be
labeled. The operands used in the equations are the output of the
delaying/shaping elements (named from \verb+i0+ to \verb+i9+). While each
equation output is associated to an activity counter, only the first four
(named \verb+S0+ to \verb+S3+) are connected to an output. As detailed in
section~\ref{SoftSec}, any combinatorial equation can be entered. Trigger
equation examples can be seen in fig.~\ref{GUI_UCTM_NB}.

\end{enumerate}

Additionally, for selecting whether activity rate are presented instead of
absolute counts, a periodic reset (every second) of the counters can be
requested by checking the checkbox located at the bottom left of the panel, 


Finally, the last setting is to choose before starting a run whether the
counting and trigger generation must be free running or set to last only for a
predefined amount a time.

\section{Example applications}
\label{ExampleApplications}

As an illustration of the capabilities of the developed unit, we describe the
utilization of the board for two simple lab experiments.

\subsection{Measurement of the muon lifetime}

\begin{figure}[ht]
\begin{center}
\includegraphics[width=0.8\textwidth]{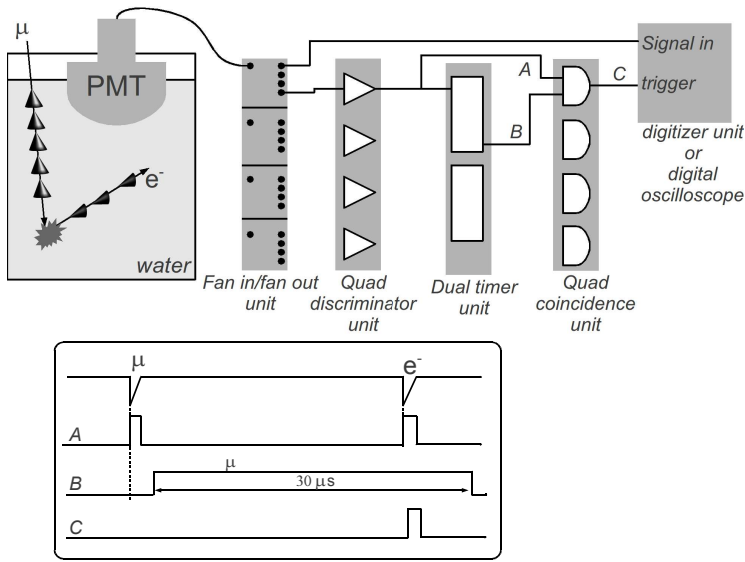}
\caption{Sketch of the lab experiment used to estimate the lifetime of cosmic muons in water. See text for details.}
\label{MuonLifeTime}
\end{center}
\end{figure}

The goal of this lab experiment is to measure the lifetime of the muon particle
using cosmic muons from atmospheric shower \cite{Hall}. The detector (see
fig.~\ref{MuonLifeTime}) is made of a simple tank filled with water having its inner surface covered with Tyvek\textregistered \cite{tyvek}. A
photomultiplier tube (PMT) is placed at the top of the tank with its window immersed in
the water. Muons crossing the tank produce Cerenkov light which is reflected by
the Tyvek\textregistered\  lining and is then detected by the PMT. Due to ionization, muons are loosing
energy in the water and the muons stopped in the tank are decaying in an electron
and two neutrinos. The electron is typically produced with a velocity above the
Cerenkov threshold and therefore also produces Cerenkov light.  Consequently the
expected signal from a muon decaying in the tank is a succession of two
Cerenkov pulses.  
In this experiment, the methodology used is to trigger on such events
and to digitize the corresponding signal using a data acquisition board or a digital
oscilloscope. The time interval between the two pulses is then measured off line.  The
classical way of generating this trigger signal is to use several electronics NIM units
(see fig.~\ref{MuonLifeTime}):
\begin{itemize}
\item A leading edge discriminator to detect the Cerenkov pulse, the output is a logic gate (A),
\item A timer unit to produce a 30\,\textmu s width gate (B) trigged by (A),
\item A logic unit to trigger on the coincidences between (A) and (B),
\item A scaler unit to count the number of coincidences.
\end{itemize}

\begin{figure}[ht]
\begin{center}
\includegraphics[width=0.9\textwidth]{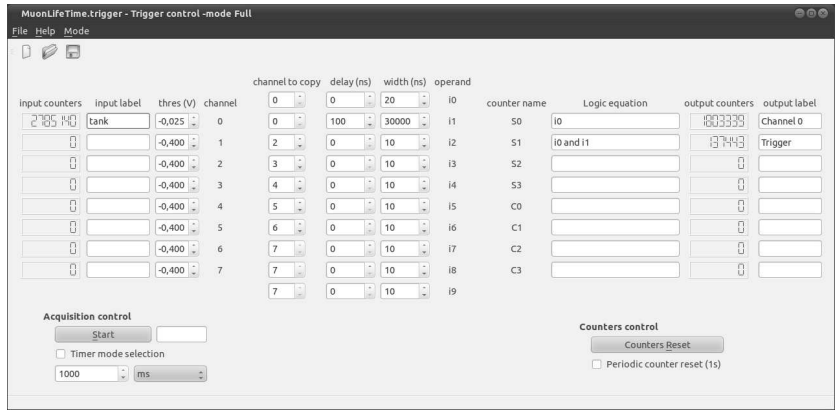}
\caption{Screen capture of the GUI configured to run the muon lifetime experiment.
\label{GUI_MuonLifeTime}}
\end{center}
\end{figure}

The trigger unit presented in this paper can be used to replace all the above
NIM units. Fig.~\ref{GUI_MuonLifeTime} shows the configuration enter in the GUI
to perform this measurement. The PMT anode is connected to the input~0 of the
board. The threshold value is chosen as -25\,mV to have a good detection
efficiency. The logic signal is duplicated to form one short pulse having a
width of 20\,ns and a large pulse of 30\,\textmu s delayed by 100\,ns to avoid
auto coincidence. The equation (S1) is a logical AND between the two pulses.
The corresponding output is connected to the digital oscilloscope to trigger
the acquisition.            

\subsection{Measurement of the atmospheric muon flux}
\begin{figure}[ht]
\begin{center}
\includegraphics[width=0.8\textwidth]{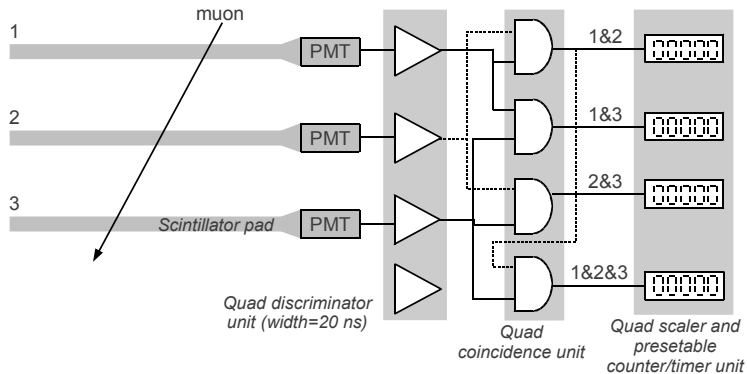}
\caption{Sketch of the lab experiment used to estimate the cosmic muon flux. See text for details.}
\label{MuonFlux}
\end{center}
\end{figure}

To measure the flux of cosmic ray muons in the lab, we use a set of three
plastic scintillator paddles placed one below the other (see
fig.~\ref{MuonFlux}). Each paddle is optically connected to a PMT with a
light guide. The use of three plastic detectors allows the
simultaneous measurement of four independent quantities
that are the number of coincidences occurring during a counting time T and corresponding to the following:
\begin{itemize}
 \item $N_{12}$: coincidences between paddles 1 and 2,
 \item $N_{13}$: coincidences between paddles 1 and 3,
 \item $N_{23}$: coincidences between paddles 2 and 3,
 \item $N_{123}$: coincidences between paddles 1, 2 and 3.
\end{itemize}
These four measurements can be used to estimate the four unknown quantities
$\epsilon_1,\epsilon_2,\epsilon_3$ and $\phi$ where $\epsilon_i$ is the
detection efficiency for the channel $i$ and $\phi$ is the muon flux.
They are related to the measurement by the equations:
\begin{equation*}
N_{ij} = T \phi A_{ij} \epsilon_i \epsilon_j
\end{equation*}
for $i \ne j$ and
\begin{equation*}
N_{123} = T \phi A_{13} \epsilon_1 \epsilon_2 \epsilon_3
\end{equation*}
where $A_{ij}$ are the geometric acceptances for the paddles $ij$ that can be
easily computed using Monte Carlo integration.

\begin{figure}[ht]
\begin{center}
\includegraphics[width=0.9\textwidth]{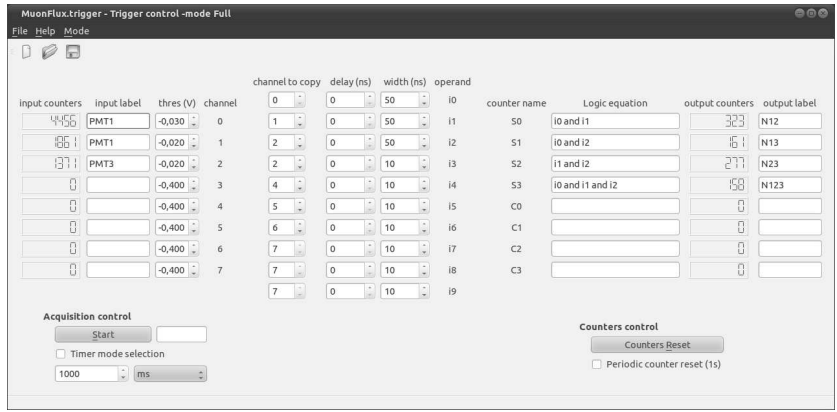}
\caption{Screen capture of the GUI configured to measure the atmospheric muon flux.
\label{GUI_MuonFlux}}
\end{center}
\end{figure}

Using standard NIM electronics, the measurement of the four quantities
$N_{12}$, $N_{13}$, $N_{23}$ and $N_{123}$ typically requires a four channel
discriminator unit, a four channels coincidence unit and a four channels scaler (see
fig.~\ref{MuonFlux}).  Fig.~\ref{GUI_MuonFlux} shows the GUI configuration to
perform the same measurement with the UCTM unit. In this example, the three PMT
anodes are connected to the first three input channels of the unit. A threshold
of -20\,mV is used for the discriminator. The discriminator outputs are not
delayed and have a fixed width of 50\,ns.  Finally, the four equations (\verb+S0-S3+)
are used to build the different coincidence signals which are connected to the
$N_{12}$, $N_{13}$, $N_{23}$ and $N_{123}$ labeled counters.

\section{Summary}
\label{SummarySec}
A configurable trigger scaler and delay NIM module has been designed to equip
nuclear physics experiments and lab teaching class. It is configurable through
a Graphical User Interface (GUI) and provides a large number of possible
trigger conditions without any Hardware Description Language (HDL) knowledge.

The module has eight discriminator inputs with individually configurable
thresholds. The discriminated version of the inputs can be logically duplicated
and used in trigger equations that are entered as plain string in the control
and readout GUI. Out of the eight possible trigger equations, four are used as
duplicated trigger outputs in the NIM standard. Scalers are available on each
input and output.

Possible future work includes: upgrading the FPGA to a faster and up to date
version in order to gain in latency and to reduce the jitter. Also with newer
FPGA, it will be possible to implement Time to Digital Conversion (TDC)
functions. Moreover dead time and live time counters will be added for each logical equation.
The possibility to manage positive inputs and to feature windowed
discriminators is also envisioned. Finally, for long term usage, it may also be
desirable to be able to pre-configure a board for a given functionality in order
to permanently replace a NIM module. A LCD display would then be added on the
front panel to report the module number and configuration file used.


\end{document}